\documentclass{article}
\usepackage{graphicx}
\begin{document}
\begin{center}
{\Large \bf Nuclear re-interaction effects in quasi-elastic
       neutrino nucleus scattering}

\vspace{1.5cm}
{\large G. Co\,', C. Bleve, I. De Mitri, and D. Martello} \\ 
\vspace{1.cm}
{Dipartimento di Fisica,  Universit\`a di Lecce, \\
 and \\
Istituto Nazionale di Fisica Nucleare  sez. di Lecce, 
\\ I-73100 Lecce, Italy}
\end{center}
\vskip 1.5 cm 

\begin{abstract}
  The quasi-elastic $\nu$-nucleus cross section has been calculated by
  using a Fermi gas model corrected to consider the re-scattering
  between the emitted nucleon and the rest nucleus. As an example of
  the relevance of this effect we show results for the muon production
  cross section on $^{16}$O target.  
\vspace{1pc}
\end{abstract}
The construction and the planning of new experiments with the
objective of detecting neutrinos, have raised great attention on the
$\nu$-nucleus interaction.  From the nuclear physics point of view,
new $\nu$-nucleus data offer the opportunity to further test the
knowledge of the nuclear structure and, perhaps, to reveal new nuclear
effects.  From the astrophysics point of view, a better determination
of the $\nu$-nucleus cross section would help to improve the
understanding of many phenomena, like star burning, element
production, supernova explosion and cooling.

Also from the point of view of an elementary particle physicist,
interested in revealing the properties of the neutrinos, either coming
from natural or artificial sources, the knowledge of the $\nu$-nucleus
interaction is important. The reaction between neutrinos and nuclei 
is the basic physical process upon which many detectors are built.
The understanding of their sensitivity to the neutrinos properties   
and the evaluation of the neutrino fluxes strongly depend
on a precise knowledge of the $\nu$-nucleus cross section.

Many Monte Carlo codes simulating detector responses, describe the
$\nu$-nucleus cross section by folding the free $\nu$-nucleon cross
section with a Fermi gas distribution. In this model, due to Smith and
Moniz \cite{smi72}, the effects of the Fermi motion and of the Pauli
exclusion principle are taken into account.  On the other hand, many
other nuclear effects are neglected. The Fermi gas model considers the
nucleus as an infinite system of non-interacting nucleons. Therefore
all the phenomena related to the nuclear surface and to the
interaction between nucleons
are not considered.  Because of its simplicity the Smith
and Moniz model is rather popular, however its intrinsic
approximations have never been thoroughly tested.

In the present contribution we analyze one of the approximations of
the Fermi gas model, specifically the one neglecting the
re-interaction between the emitted nucleon and rest nucleus. We have
developed a model to consider this effect. Details can be found in
ref. \cite{ble01}.  We present here results for the muon production on
the $^{16}$O nucleus. They have been calculated for the kinematic
conditions suggested as benchmark by the conference organizers.

We have already remarked that the range of applicability of the Fermi
gas model in treating nuclear excitations is restricted to situations
where the surface effects can be neglected. For this reason the Fermi
gas model can be reliably used only in the quasi-elastic region.
Roughly speaking, this region can be identified with a range
of nuclear excitation energies from 50 to  300 MeV.

%
%
%
\begin{figure}
\begin{center}
\includegraphics*[scale=0.9]
{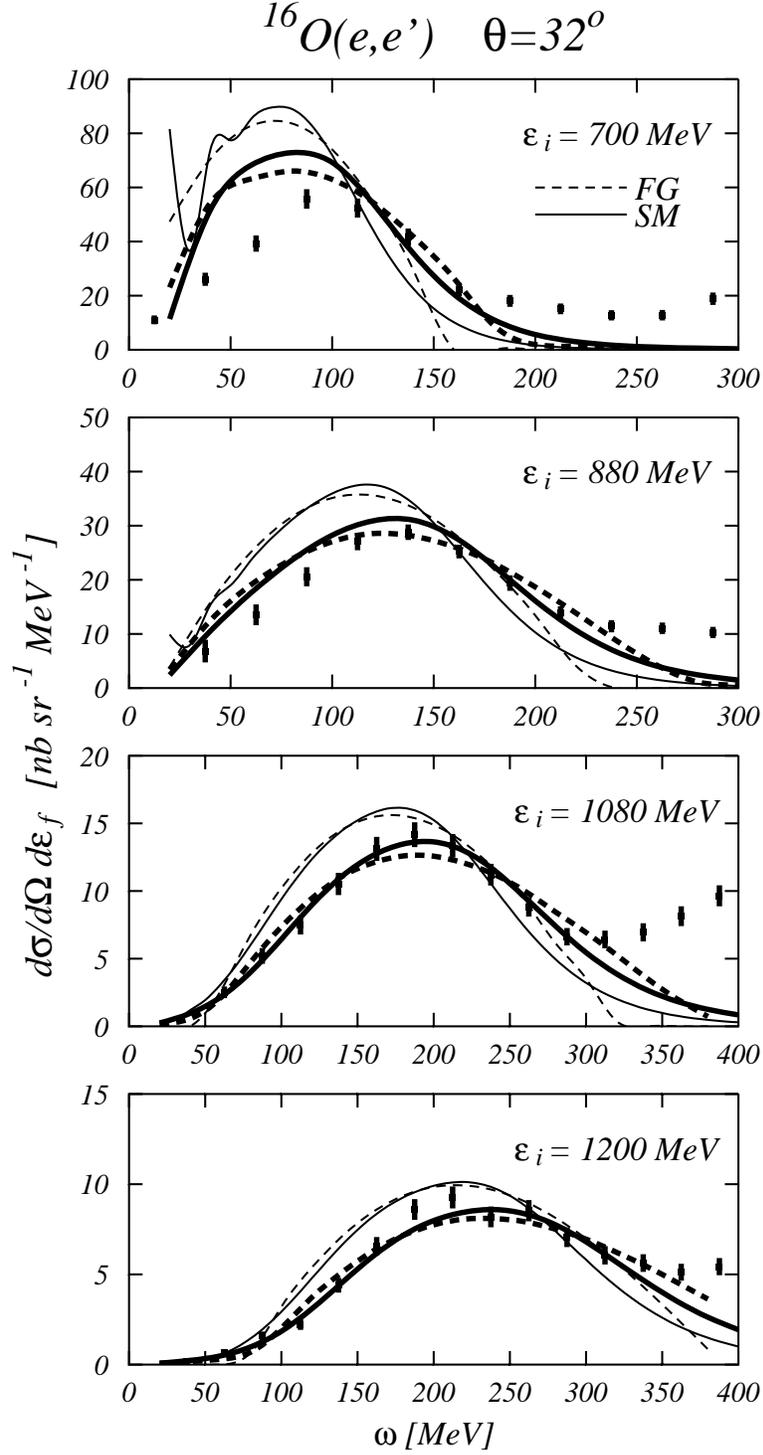}
\caption{\small  
  Comparison between Fermi gas (FG) and Continuum Shell model (SM)
  results. We have indicated with $\epsilon_i$ the initial electron
  energy, and with $\omega$ the missing energy, corresponding in our
  model with the nuclear excitation energy.  
  The lower lines have been obtained by including the Final State
  Interaction.
  Data from  ref. \protect\cite{ang96}.
}
\label{fig:ee}
\end{center}
\end{figure}

The validity of the infinite system approximation in the description
of the quasi-elastic excitation has been studied in ref. \cite{ama94}
by comparing electromagnetic responses calculated within the Fermi gas
model and the Continuum Shell model. From the physics contents these
two models differ only because the Continuum Shell model considers the
finite nuclear dimensions.  In ref. \cite{ama94} it is shown that a good
agreement between the Fermi gas and the Continuum Shell model results 
can be obtained
when
an appropriate value of the Fermi momentum is used. It is also shown
that this is a much better prescription than the commonly used Local
Density Approximation.

The thin lines of fig. \ref{fig:ee} show the electron scattering cross
sections on the $^{16}$O nucleus calculated with the two different
models.  We used the value of the Fermi momentum of 216 MeV/c obtained
with the average momentum prescription presented in ref. \cite{ama94}.
The agreement between the results of the two calculations improves
with increasing the energy of the electron.

In the figure, the agreement of the thiner lines with the experimental
data \cite{ang96} is rather poor.  This is a common feature of
all the calculations done in the quasi-elastic region with nuclear
models neglecting the re-interaction between the emitted nucleon and
the rest nucleus \cite{ama01}.  To include this effect, commonly called
Final State Interaction (FSI), we used the model developed in ref.
\cite{co88}.

In this model the response containing the FSI can be expressed in
terms of folding integral of the response without FSI:
\begin{eqnarray}
&~&
\nonumber
S^{FSI}(|{\bf q}| , \omega)  = \\
&~&
\int_0^\infty dE \,\,\,
S^0(|{\bf q}| , E) 
\left[ \rho(E,\omega)+\rho(E,-\omega)
\right]
\label{eq:res}
\end{eqnarray}
where the folding function is:
\begin{equation}
\rho(E,\omega)= 
\frac{1}{2\pi}
\frac{\Gamma(\omega)}
{ 
[ E-\omega-\Delta(\omega) ]^2 +
[ \Gamma(\omega)/2]^2 
}
\label{eq:rho}
\end{equation}

The two functions $\Delta$ and $\Gamma$ represent the real and
imaginary part of the so-called spreading width, which takes into
account nuclear excitations beyond those considered in mean-field
models, like the Fermi gas and the Shell models. A microscopic
evaluation of this spreading width \cite{dro87} has shown that, in the
quasi-elastic region, its effects can be rather well described in
terms of an optical potential. For this reason, as it is commonly 
done \cite{ama01,co88}, we fixed the $\Delta$ and $\Gamma$
functions as volume integrals of optical potentials reproducing
elastic nucleon-nucleus cross sections data.
%
%
\begin{figure}
\begin{center}
\includegraphics[scale=0.7]
{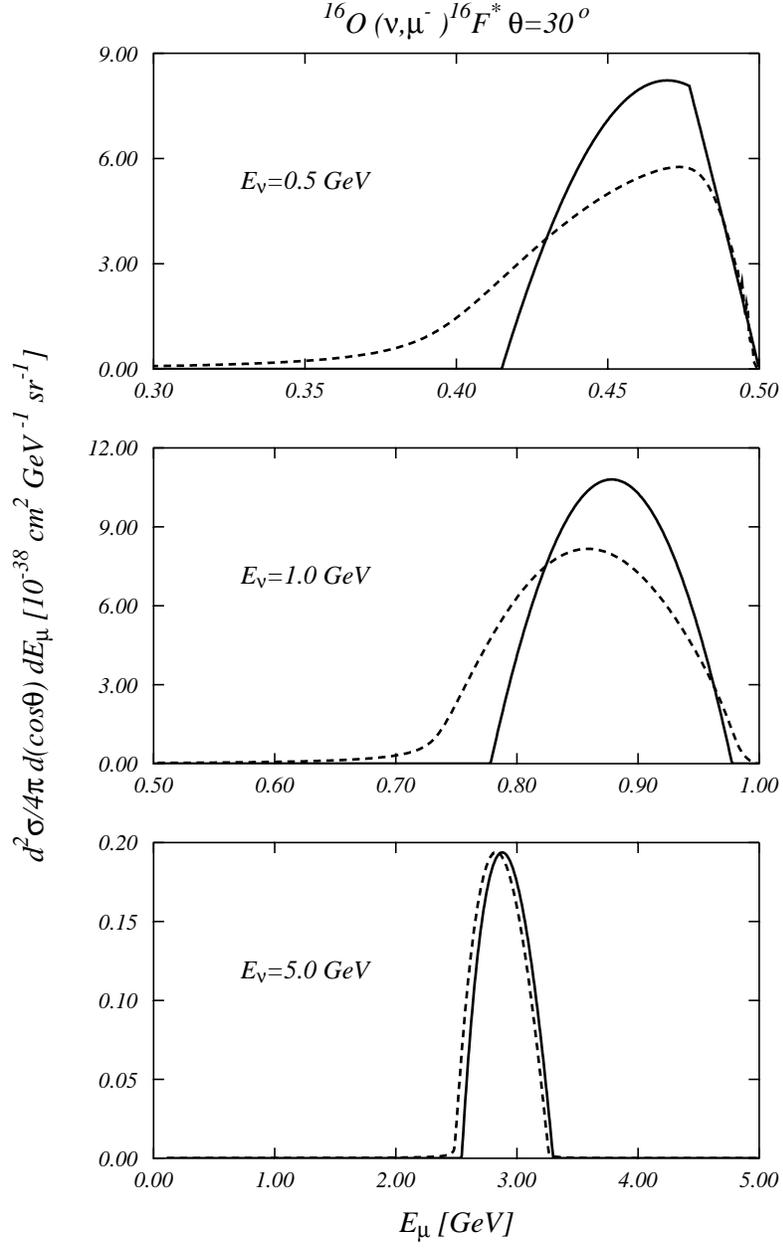}
\caption{\small  
  Double differential $\nu$-nucleus cross section as a function of the
  emitted muon energy for three values of the neutrino energy.  The
  full lines have been calculated with the Fermi gas model and the
  dashed lines include the FSI effects.  
  }
\label{fig:x2}
\end{center}
\end{figure}

The thick curves of fig. \ref{fig:ee} have been obtained from the thin
lines by applying the procedure above outlined.  The maxima of the
mean field responses are lowered, and part of the strength is shifted
towards higher excitation energies. The shift of the position of the
maxima is due to the presence, in our model, of an effective nucleon
mass smaller than the bare nucleon mass. This effective mass takes into
account the non locality of the nuclear mean field.

It is evident from the figure, that the inclusion of the FSI largely
improves the agreement with the data.  The features we have discussed
are not typical of the specific example considered. The quasi-elastic
electron scattering data can be reproduced only when the FSI is
considered \cite{ama01}.
%
%
\begin{figure}
\begin{center}
  \includegraphics[scale=0.5] {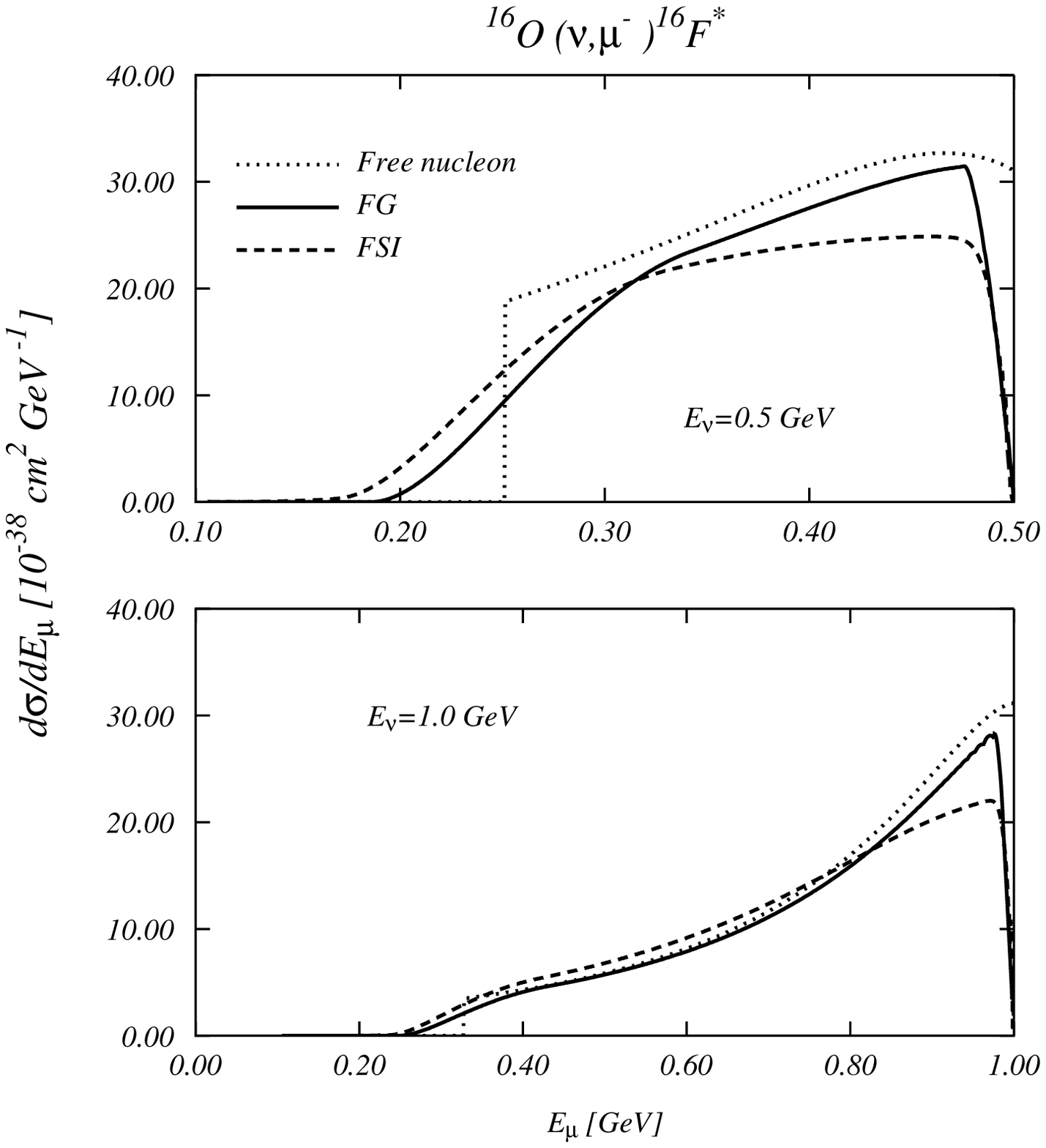}
\caption{\small  
  Cross section integrated on the muon emission angle as a function of
  the muon energy.
}
\label{fig:xene}
\end{center}
\end{figure}

From the above considerations it appears evident that a realistic
description of the $\nu$-nucleus interaction in the quasi-elastic
region requires the treatment of the FSI. Our calculations of the
$\nu$-nucleus cross section have been done by using the FSI model
above described \cite{ble01}. As eq. (\ref{eq:res}) indicates, we need
a mean field response. We have used the Fermi gas response of the
Smith and Moniz approach implemented as in ref. \cite{lip95}.

The double differential cross section for the $^{16}$O
($\nu_{\mu},\mu$) $^{16}$F reaction is shown in fig. \ref{fig:x2} as a
function of the emitted muon energy for three values of the neutrino
energy. The emission angle of the muon has been fixed at 30$^0$. The
full lines show the result obtained with a bare Fermi gas
calculations, while the dashed lines contain the effects of the FSI.

A first observation is that for E$_\nu$=5.0 GeV the FSI
effects are negligible. The only difference between the two curves is
a small shift of the position of the peak 
due to the presence of the effective nucleon mass. 
For this reason henceforth we shall neglect the results
of the  5.0 GeV calculations.

The FSI effects are instead rather important for the other two
energy values. At 0.5 GeV the Fermi gas response still shows the
linear behaviour at high muon energies (small nuclear excitation
energies). This is a signature of the fact that, under these kinematic
conditions, some of the particle-hole transitions are Pauli blocked.
The linear behaviour is no longer in the 1 GeV neutrinos
cross section.

In both cases the FSI strongly modifies the responses, decreasing
their values in the peak region and shifting strength in the forbidden
region. The FSI increases the cross section at small muon energies and
decreases it at higher energies.

These modifications of the nuclear response have consequences on 
the integrated cross sections. In fig. \ref{fig:xene} we
show the cross sections integrated with respect to the muon emission
angle, as a function of the muon energy.  Together with the Fermi gas
and FSI results we also show the free nucleon cross section.

%
%
%
\begin{figure}
\begin{center}
\includegraphics[scale=0.8] {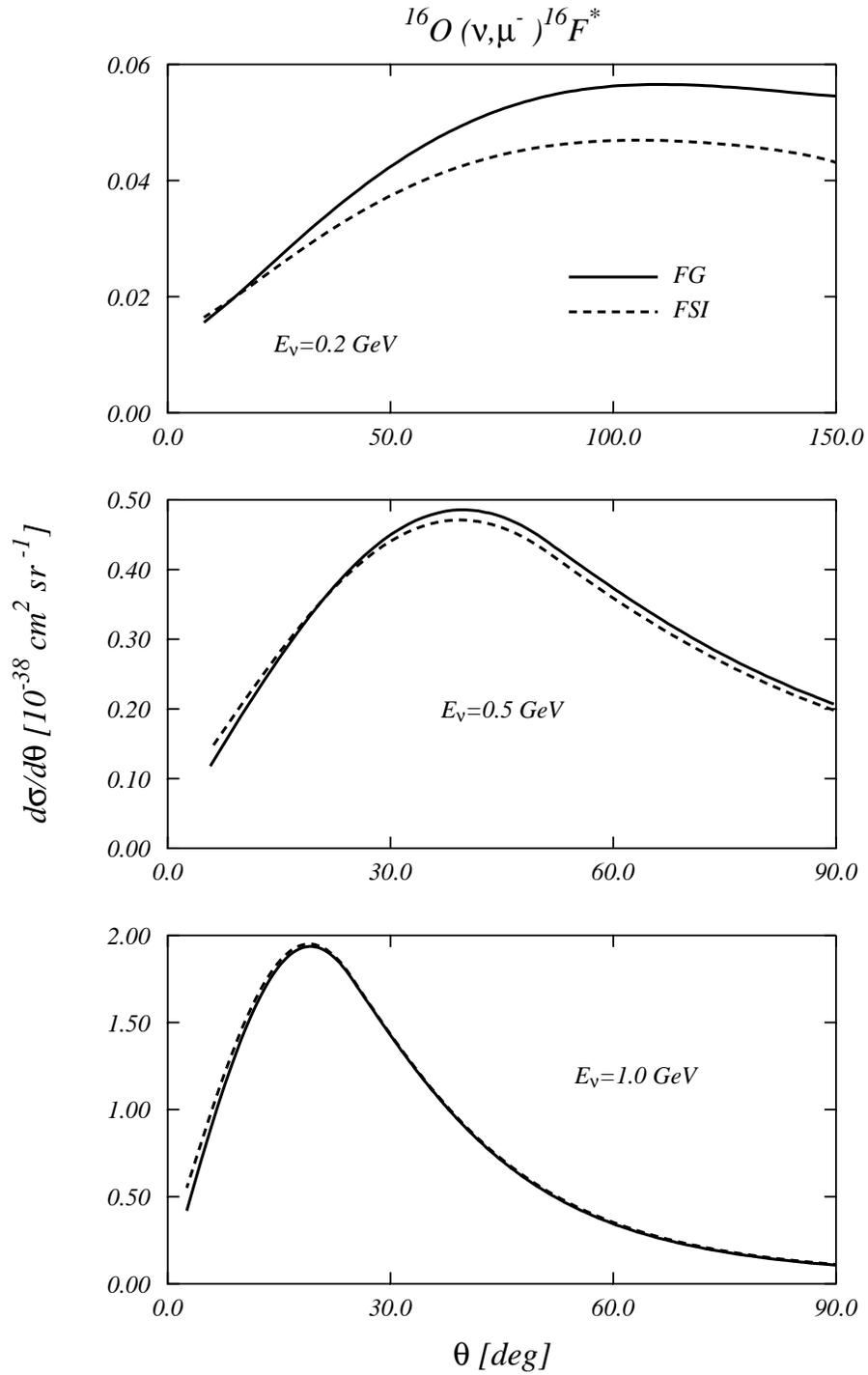}
\caption{\small  
 Cross sections integrated on the energy of the emitted muon as a
 function of the scattering angle.
}
\label{fig:xang}
\end{center}
\end{figure}

The differences between the various calculations are more evident for
0.5 GeV case. The free nucleon cross section has a sharp energy
threshold.  When the nucleon is embedded in the nuclear medium this
sharp behaviour of the cross section is smoothed, and the cross
sections is different from zero also in sub-threshold region.  At high
values of the muon energy (low excitation energy) the effect of
the Pauli blocking is clearly visible in the nuclear cross sections.
As expected, the inclusion of the FSI lowers the Fermi gas cross
section at high muon energies and increases it at lower energies
values.  These effects become relatively smaller with increasing
neutrino energy.

In fig. \ref{fig:xang} we show the neutrino cross section
integrated over the muon energy as a function of the emission angle.
For neutrinos of 
1.0 and 0.5 GeV the curves with and without FSI are rather
similar. They are instead remarkably different for 0.2 GeV.

This result indicates that the effect of the FSI in fig. \ref{fig:x2}
is predominantly a redistribution of the strength, without loss or
increase of it.  In that figure, the integrals of the cross sections
with and without FSI are almost identical.  
Indeed if the $\Gamma$ and
$\Delta$ functions are independent of $\omega$ these integrals
would be identically the same.  
In the region for $\omega >$ 50 MeV the $\Gamma$
and $\Delta$ functions are almost constant.  For neutrino energies
above 0.5 GeV the major weight in the integral comes from the region
of constant $\Gamma$ and $\Delta$. This explain the result of the two
lower panels of fig. \ref{fig:xang}.  The situation is rather
different for E$_{\nu}$=0.2 GeV as is seen in the upper panel of
the figure.

We have shown that the FSI produces relevant modifications on the
quasi-elastic double differential $\nu_\mu$-nucleus cross section. The
main effect is a redistribution of the strength, shifted from the peak
position towards lower muon energies. The FSI effects become smaller
with increasing neutrino energy and they are negligible for 5.0 GeV
neutrinos. The inclusion of the FSI is a necessary modification of the
commonly used Fermi gas model, and more in general of any mean-field
model, in order to have a more realistic description of the neutrino
nucleus interaction.  On the other hand, if the cross section is
integrated on the energy of the emitted muon the FSI effects are
appreciable only for rather low neutrino energies (we have shown one
example of the  0.2 GeV data).

We have presented results for the muon neutrinos and the $^{16}$O
nucleus.  Our conclusions, however, are more general, because the FSI
interactions effects are intrinsic to the nuclear excitation modes,
and independent of the excitation probe. The FSI effects above
described, are present also in electron neutrino and tau neutrino
reactions and for any nuclear system.

\end{document}